\documentclass[twocolumn,aps,prl,showpacs,superscriptaddress]{revtex4}

\usepackage{graphicx,color}%
\usepackage{amsfonts}
\usepackage{amsmath}
\usepackage[colorlinks]{hyperref}

\newcommand{\id}{\mathbb{I}_2}
\newcommand{\mean}[1]{\langle#1\rangle}

\newcommand{\ket}[1]{\ensuremath{\left|{#1}\right\rangle}} 
\newcommand{\avg}[1]{\ensuremath{\left\langle{#1}\right\rangle}}
\newcommand{\sinc}[1]{\ensuremath{\textrm{sinc}\left(#1\right)}}
\newcommand{\sincq}[1]{\ensuremath{\textrm{sinc}^2\left(#1\right)}}
\DeclareMathAlphabet{\mathpzc}{OT1}{pzc}{m}{it}

\begin{document}


\title{Experimental Test of the Quantum Violation of the Noncontextuality Inequalities for the $n$-Cycle Scenario}



\author{Gilberto Borges}
\affiliation{Departamento de F\'{\i}sica, Universidade Federal de Minas Gerais,
 Caixa Postal 702, Belo Horizonte, MG 30123-920, Brazil}

\author{Marcos Carvalho}
\affiliation{Departamento de F\'{\i}sica, Universidade Federal de Minas Gerais,
 Caixa Postal 702, Belo Horizonte, MG 30123-920, Brazil}

\author{Pierre-Louis de Assis}
\affiliation{CEA-CNRS-UJF group `Nanophysique et Semiconducteurs', Institut N\'{e}el, CNRS -
Universit\'{e} Joseph Fourier, 38042, Grenoble, France}

\author{Jos\'e Ferraz}
\affiliation{Departamento de F\'{\i}sica, Universidade Federal de Minas Gerais,
 Caixa Postal 702, Belo Horizonte, MG 30123-920, Brazil}

\author{Mateus Ara\'ujo}
\affiliation{Faculty of Physics, University of Vienna, Boltzmanngasse 5, 1090 Vienna, Austria}

\author{Ad\'an Cabello}
\affiliation{Departamento de F\'{\i}sica Aplicada II, Universidad de Sevilla, E-41012 Sevilla, Spain}
\affiliation{Departamento de F\'{\i}sica, Universidade Federal de Minas Gerais,
 Caixa Postal 702, Belo Horizonte, MG 30123-920, Brazil}

\author{Marcelo Terra Cunha}
\affiliation{Departamento de Matem\'atica, Universidade Federal de Minas Gerais,
 Caixa Postal 702, Belo Horizonte, MG 30123-920, Brazil}

\author{Sebasti\~{a}o P\'adua}
\email{spadua@fisica.ufmg.br}
\affiliation{Departamento de F\'{\i}sica, Universidade Federal de Minas Gerais,
 Caixa Postal 702, Belo Horizonte, MG 30123-920, Brazil}


\date{\today}


\begin{abstract}
The inequalities that separate contextual from noncontextual correlations for the $n$-cycle scenario (consisting of $n$ dichotomic observables $\mathcal{O}_{j}$, with $j = 0, \ldots, n-1$ and such that $\mathcal{O}_{j}$ and $\mathcal{O}_{j+1}$ are jointly measurable) have been recently identified [arXiv:1206.3212 (2012)]. Here we report the results of an experiment designed to reach the maximum quantum violation of these inequalities for any even number of observables ranging from 4 to 14. The four dimensional Hilbert space required for the test was spanned by two photonic qubits encoded in the tranversal paths of photon pairs, and the joint measurability of the observables is guaranteed by measuring correlations between observables from different modes. Our results show contextual correlations as predicted by quantum mechanics.

\end{abstract}



\maketitle


{\em Introduction.---}Physical correlations are an essential resource in many technological applications. For example, a number of techniques of photon correlation spectroscopy, microscopy, and interferometry used for studying condensed matter and biological systems are based on the detection of correlated light \cite{hanbury, berland, berne, agero}. On the other hand, in quantum information, correlations between microscopic systems (photons, ions, molecules, quantum dots, etc.) have been demonstrated to be useful for secure communication and fast information processing in quantum computers \cite{chuang,gisin}.
One of the fundamental differences between these two types of correlations is that, while correlations of the first type are noncontextual (\textit{i.e.}, they can be explained assuming that the correlated results correspond to preexisting properties which are independent of which other jointly measurable observables are tested), the correlations useful for quantum information cannot be explained this way -- they are contextual. This contextuality has already been observed in various physical systems, \textit{e.g.}, ions~\cite{Kirchmair1,ZUZAWDSDK13}, photons~\cite{SZWZ00, Amselem1,DHANBSC13}, and neutrons~\cite{Bartosik1}.

The problem of finding the conditions that separate contextual from
noncontextual correlations for an arbitrary measurement scenario
(defined as a set of observables and the subsets that are jointly
measurable) is computationally intractable, and the solution is known
only for a few scenarios. This solution is given as a set of noncontextuality inequalities, that define the boundary of the set of
noncontextual correlations. The most famous sets of noncontextuality inequalities are without doubt the ones that characterize the  Clauser-Horne-Shimony-Holt~\cite{Clauser1} and the Klyachko-Can-Binicio\u{g}lu-Shumovsky~\cite{Klyachko1} scenarios, which were, respectivley, the first
nonlocality and contextuality scenarios to be completely characterized.
Recently, these scenarios have been understood as the $n=4$ and $n=5$ cases of a more general scenario, the $n$-cycle scenario (consisting of $n$ dichotomic observables $\mathcal{O}_{j}$, with $j =0, \ldots, n-1$, and such that $\mathcal{O}_{j}$ and $\mathcal{O}_{j+1}$ (with the sum modulo $n$) are jointly measurable), and the corresponding sets of inequalities have been found for arbitrary $n$~\cite{Araujo1}. This provides a valuable tool to investigate contextuality and how it evolves with the number of settings. For this specific scenario, the quantum violation occurs for all $n$, and only technical reasons make the observation of violations harder for large $n$.

In this Letter, we report the experimental observation of contextual correlations certified by violations of the tight noncontextuality inequalities described in Ref.~\cite{Araujo1} for any even number of settings ranging from $n=4$ to $n=14$. For each $n$, the experiment is designed to fulfill all the requirements for a test of contextuality and aim for the maximum quantum violation. 


{\em Methods.---}To certify contextuality we use the violation of the noncontextuality inequalities \cite{Araujo1},
\begin{equation}
 \label{eq:ncinequality}
 \Omega = \sum_{j=0}^{n-2} \mean{\mathcal{O}_j\mathcal{O}_{j+1}}
 -\mean{\mathcal{O}_{n-1}\mathcal{O}_0} \stackrel{\text{\tiny{NCHV}}}{\leq} n-2,
\end{equation}
for even $n \ge 4$, where $ \stackrel{\text{\tiny{NCHV}}}{\leq} n-2$ indicates that $n-2$ is the highest value allowed for noncontextual correlations.

Any experimental test of a noncontextuality inequality should satisfy two conditions. One is that the observables whose correlations are considered in the inequality should be jointly measurable \cite{GKCLKZGR10}. For inequality \eqref{eq:ncinequality}, this means that $\mathcal{O}_{j}$ and $\mathcal{O}_{j+1}$ should be jointly measurable for any $j=0,\ldots,n-1$. The second requirement is that every observable $\mathcal{O}_j$ has to be measured using the same experimental configuration in every context \cite{ABBCGKLW12}. For testing inequality \eqref{eq:ncinequality}, this means that, for any $j=0,\ldots,n-1$, the experimental configuration used for measuring $\mathcal{O}_j$ should be the same both when $\mathcal{O}_j$ is jointly measured with $\mathcal{O}_{j+1}$ and when $\mathcal{O}_j$ is jointly measured with $\mathcal{O}_{j-1}$.

To assure that these two requirements are achieved in our experiments, we use a two-photon system in which each photon encodes a qubit and is initially prepared in the state
\begin{equation} \label{phi+_methods}
 \ket{\phi^+} = \frac{1}{\sqrt{2}}(\ket{00}+\ket{11}),
\end{equation}
{and implemented the observables:}
\begin{equation}
	\mathcal{O}_j =
	\begin{cases}
 O(\theta_j) \otimes \id & \text{ for even }j,\\
 \id \otimes O(-\theta_j) & \text{ for odd }j,
 \end{cases}
 \label{eq:observables}
\end{equation}
{where $\id$ is the identity in the Hilbert space of one qubit.} 

\begin{figure}[htpb]
\centering
 \includegraphics[width=8.5cm]{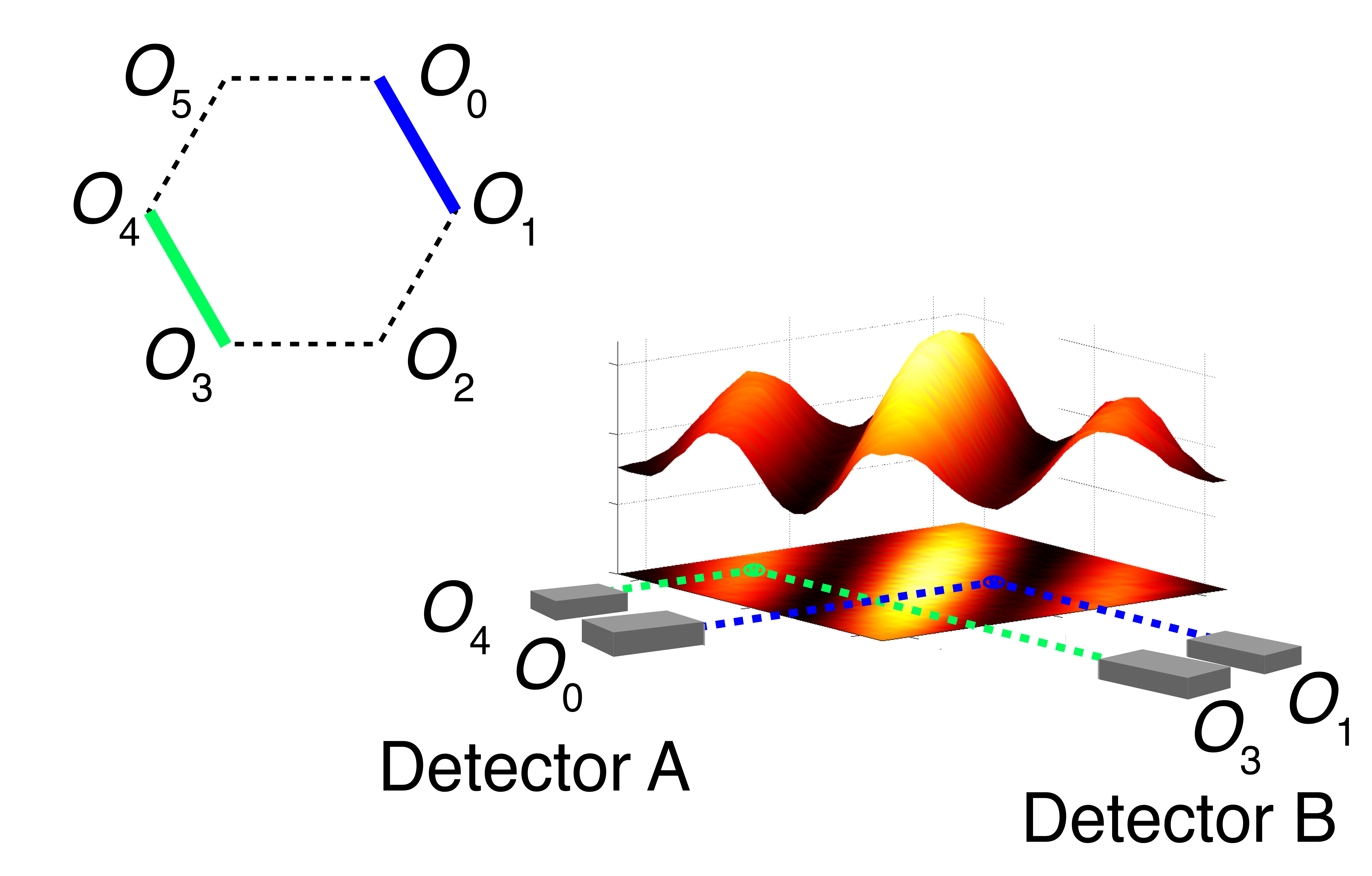}
\caption{ (Color online) Left up corner: Graph of compatibility for the $n=6$ scenario. Vertices, $O_{i}$ represent dichotomic observables and adjacent vertices represent jointly measurable observables. Right: An example of the experimental implementation of two correlations, $O_0O_1$, in blue, and $O_3O_4$, in green. In our experiment, different detector positions correspond to the implementation of different observables.}
\label{Fig1}
\end{figure}

The fact that the measurements of $\mathcal{O}_j$ and $\mathcal{O}_{j+1}$ are always performed on different particles ensures joint measurability. Measuring $\mathcal{O}_{j}$ together with $\mathcal{O}_{j+1}$ or with $\mathcal{O}_{j-1}$ does not require any change in the experimental configuration used for measuring $\mathcal{O}_{j}$, so we can guarantee that every observable is measured the same way in every context (Fig.~\ref{Fig1}).

{For the prepared state (Eq.~\eqref{phi+_methods}) the maximal quantum value of $\Omega$, is attained for the observables:}
 \begin{equation}
O(\theta_j) = \cos(j \pi/n) \sigma_x +\sin(j \pi/n) \sigma_y,
\end{equation}
with $\sigma_x$ and $\sigma_y$  the Pauli matrices $x$ and $y$.


{\em Experimental implementation.---}State $\ket{\phi^+}$ can be produced using a double-slit (DS) to encode two qubits in the transversal path of photon pairs generated by spontaneous parametric down conversion (SPDC)~\cite{Neves1, Neves2}. In the near field of the double-slit aperture, the quantum state is described by~\cite{Peeters1, deAssis1}
\begin{equation}
\ket{\phi^+} = \frac{1}{\sqrt{2}}\left\{\ket{0}_{s}\ket{0}_{i} + \ket{1}_{s}\ket{1}_{i}\right\} \!,
\label{phi+}
\end{equation}
where $s(i)$ refers to signal (idler) photons and $\ket{j}$ is the state of the photon that crossed the slit $j$ ($j = 0,1$).

\begin{figure}[htpb]
\centering
\includegraphics[width=8.5cm]{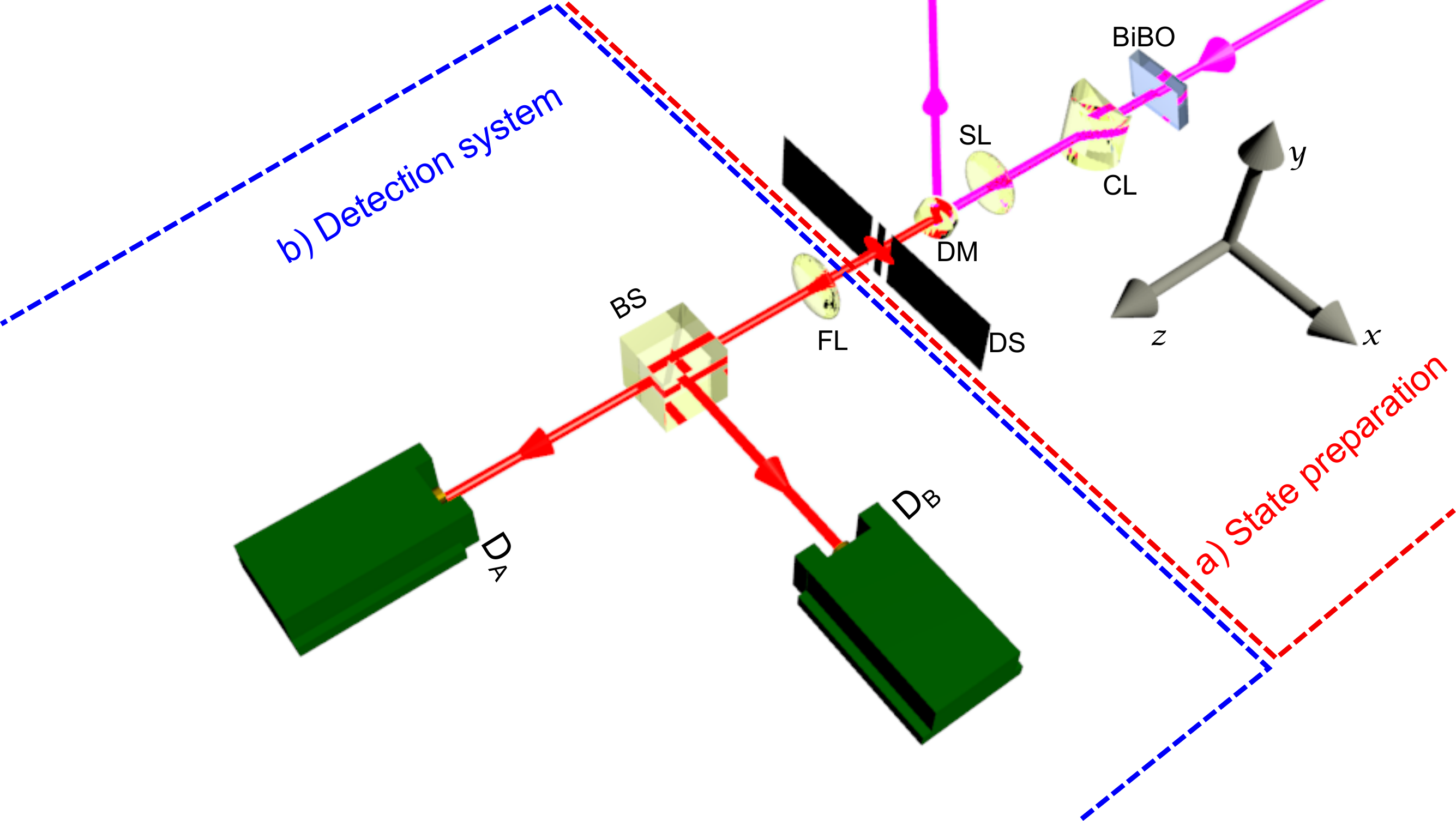}
\caption{(Color online) Experimental setup. (a) State preparation. A CW laser generates colinear photon pairs in a BIBO crystal. A non confocal telescope, consisting of a spherical lens SL$_{2}$ and of a cylindrical lens CL, projects the magnified image of the crystal in the DS plane $x$ direction. Dichroic mirror DM reflects the pump beam. (b) Detection system. The lens FL is used to project the DS far field in the detection plane. BS is a 50/50 beam splitter and D1 and D2 are APD's detectors.}
\label{expsetup}
\end{figure}

In Fig.~\ref{expsetup}, we show a schematic view of our experimental apparatus which is divided in two parts. The first being the state preparation and the second is the detection setup.


For the state preparation, a continuous diode laser with exit power of 50 mW, operating at 405 nm, is focused in the center of a 2 mm $\textrm{Bi}\textrm{B}_3\textrm{O}_6$ crystal (BIBO) by a spherical lens (not shown on Fig.~\ref{expsetup}), f$_{1}$ = 30 cm, generating collinear type I SPDC pairs of photons. The photon pairs ($\lambda$ = 810 nm) and the laser beam propagate along the $z$ direction and, immediately after the crystal, a dichroic mirror (DM) reflects the pump beam and transmits the down-converted photons.
The double-slit aperture is placed 40 cm distant from the BIBO crystal, perpendicular to the $z$ direction, each slit of the DS has width 2a = 80 $\mu$m, along the $x$ direction, the center to center separation is $d = 160$ $\mu$m. Between the crystal and the DS, a linear optical setup controls the quantum correlations of the twin-photons in the aperture plane \cite{Peeters1}. This setup is composed by a cylindrical lens (CL) of focal length f$_{CL}$ = 5.0 cm and a spherical lens (SL$_{2}$) with focus f$_{2}$ = 20 cm. This scheme projects a magnified image of the crystal center onto the DS plane in the $x$ direction. With this experimental configuration, we are able to control the correlations of the down-converted photons, such that the photon pair always passes through the same slit of the DS, thus forming the entangled state represented in Eq.~(\ref{phi+}). The length of the DS larger dimension along the $y$ direction is 8.0 mm, larger than the down-converted beam width and it can be considered infinite.

The detection system is set to project the Fourier transform of the DS at the detection plane, as shown in Fig.~\ref{expsetup}(b). In this scheme, we use a spherical lens (FL) of focal length 30 cm in the $f - f$ configuration, and two avalanche photo-diodes (APD) detectors at the exit port of a balanced beam splitter (BS). The detectors are placed 60 cm from the DS plane and equipped with interference filters, centered at 810 nm (10 nm FWHM bandwidth). In front of each detector we have a pinhole with diameter $2b = 200$ $\mu$m, for spatial filtering. The APD's detectors are mounted in translation stages and can be scanned in the $x$ direction. Coincidences between the detectors are obtained with a homemade electronic circuit with 5.4 ns of temporal window.
A two dimensional $25 \times 25$ array of coincidence counts was obtained by scanning both detectors in the $x$ direction of the far field plane, with step length 100 $\mu$m and acquisition time of 30 s for each point. 


{\em Experimental results and discussion.---}In order to test experimentally the violation of the inequality~\eqref{eq:ncinequality}, we first need to find the correspondence between the coincidence counts and our observables, defined in Eq.~\eqref{eq:observables}, \textit{i.e.}, we need to know how the measurement operator is implemented when the coincidence counts is acquired in the DS far field.

The mathematical description of the measurement operator in the transversal direction is
\begin{equation}
\Pi\left(x_i,x_s\right)= \Pi_s\left(x_s\right)\otimes \Pi_i\left(x_i\right) \!,
\label{Pi}
\end{equation}
where each single system operation $\Pi_\nu\left(x_\nu\right)$ ($\nu=i, s$) is given by:
\begin{align}
\Pi_\nu\left(x_\nu\right) = & \mathcal{A}_\nu\left(x_\nu\right)\left\{\mathbb{I}_{2} + \sinc{\kappa\,b} \times \right. \nonumber \\
 & \left. \times \left[\cos\left(\kappa x_{\nu} \right)\sigma_x^\nu+\sin\left(\kappa x_{\nu} \right)\sigma_y^\nu\right]\right\} \!,
\label{qbit_op}
\end{align}
$\mathbb{I}_{2}$ is identity operator, $\kappa=k_{p}d/2f_{FL}$, k$_{p}$ is the pump beam wavenumber, 2$b$ is the transversal dimension of the detectors, and $x_\nu$ ($\nu =i, s$) is the detector $\nu$ transversal position. The Pauli matrices $\sigma_x^\nu$ and $\sigma_y^\nu$ in Eq.~(\ref{qbit_op}) are written in terms of the slit states $\ket{j}$ ($j = 0,1$).  The functions $\mathcal{A}_{\nu}\left(x_{\nu}\right)$ are given by
\begin{equation}
\mathcal{A}_\nu\left(x_\nu\right)=\frac{k_pa}{2 \pi f_{FL}}\sincq{\frac{\kappa x_{\nu}}{d}} \!,
\label{dif_env}
\end{equation}
which is the diffraction envelope of the interference pattern. The measurement operator is physically implemented by the FL spherical lens in the $f-f$ configuration and the APD detectors at the FL focal plane, see Fig.~\ref{expsetup}(b). The measurement operator $\Pi_\nu\left(x_\nu\right)$ is equal to
$E_\nu^{-}\left(x_\nu\right)E_\nu^{+}\left(x_\nu\right)$, with $E_\nu^{+}\left(x_\nu\right)$, resp.~$E_\nu^{-}\left(x_\nu\right)$, being the positive (negative) frequency electric field operator at the detection plane, and is written in terms of the slit states $\ket{j}$ ($j = 0,1$) \cite{Neves1,Neves2,Neves3}.

The expected value of operator $\Pi\left( x_i, x_s\right)$ (see Eq.~\eqref{Pi}) is proportional to the coincidence counts at the detector positions $x_i$ and $x_s$.
The connection between the mean value of the jointly measurable dichotomic observables and the experimental coincidence counts is given by
\begin{widetext}
\begin{equation}
\avg{O_i(\theta_i) \otimes O_s(\theta_s)}=\frac{\mathpzc{C}\left(\theta_i,\theta_s\right)+\mathpzc{C}\left(\theta_i-\pi,\theta_s-\pi\right)-\mathpzc{C}\left(\theta_i-\pi,\theta_s\right)-\mathpzc{C}\left(\theta_i,\theta_s-\pi\right)}{\mathpzc{C}\left(\theta_i,\theta_s\right)+\mathpzc{C}\left(\theta_i-\pi,\theta_s-\pi\right)+\mathpzc{C}\left(\theta_i-\pi,\theta_s\right)+\mathpzc{C}\left(\theta_i,\theta_s-\pi\right)} \!,
\label{cor_coef}
\end{equation}
\end{widetext}
where $\mathpzc{C}\left(\theta_i,\theta_s\right)$ is the coincidence count associated with the $\theta_{i}$ and $\theta_{s}$ interference pattern angular position, and $O_\nu$ ($\nu = i,s$) are given by
\begin{equation}
O_\nu=\cos\left(\theta_\nu \right)\sigma_x+\sin\left(\theta_\nu \right)\sigma_y \:.
\label{os}
\end{equation}
Eq.~(\ref{cor_coef}) is general, for any pair of angles $\theta_i$ and $\theta_s $. 
The angular settings of Eq.~(\ref{eq:observables}) and Eq.~(\ref{cor_coef})  are related with the detector positions by
\begin{equation}
\theta_\nu = \kappa x_\nu \:,
\label{ang_pos_rel}
\end{equation}
which explains how each pair of detector position relates to each term in the Inequality~(\ref{eq:ncinequality}), see again Fig.~\ref{Fig1}.

The violation of the noncontextuality inequalities for the $n$-cycle is achieved when we use ${n}/{2}$ different positions for each detector, registering the coincidence counts  to calculate the correlations coefficients of Eq.~(\ref{cor_coef}). To better explain the procedure utilized, let us use the $n = 4$ (CHSH) scenario as an example. In this case, an optimal setting for the signal subsystem measurement apparatus is: $\left\lbrace\theta_s\right\rbrace = \left\lbrace0, \pi/2\right\rbrace$ and for the idler: $\left\lbrace\theta_i\right\rbrace =\left\lbrace\pi/4, 3\pi/4\right\rbrace$. 
To obtain the correlation coefficients it is also necessary to use the orthogonal measurement operators, which are obtained when we change the subsystem angles by a factor of $\pi$, \textit{i.e.}, $\left\lbrace\theta_s - \pi\right\rbrace = \left\lbrace-\pi, -\pi/2\right\rbrace$ and $\left\lbrace\theta_i - \pi\right\rbrace = \left\lbrace-3\pi/4, -\pi/4\right\rbrace$. 
Since the scanning process does not generate all the optimal points for maximal quantum violation, we used the nearest available experimental data. This problem reduces the amount of violation, but not dramatically.
As we increase the value of $n$, the above procedure is repeated, until we reach the $n = 14$ value, above which the angular separation between two consecutive observables in the same subsystem is smaller than the experimental resolution of the position.

In Table~\ref{tab1}, we report our experimental results for each $n$. The Table shows three different values of $\Omega$. The first one is obtained using the raw experimental coincidence counts, denoted as $\Omega_{\rm exp}$. The second one, denoted as $\Omega_{\rm exp}^{\rm bd}$, is the value predicted by quantum mechanics assuming ideal equipment when we consider that the angular separation between two operators that define a context is $\gamma_{\rm exp} = \kappa(x_i - x_s)$ instead of $\gamma = \pi/n$. In other words, this value of $\Omega$ is the one obtained when we use  quantum mechanics, the angles experimentally implemented, and the maximally entangled state defined by Eq.\eqref{phi+}, but no other source of imperfection (such as the preparation of a non-maximally entangled state). The third value, $\Omega_{\rm max}^{\rm bd}$, is the quantum bound using the ideal angular separation $\gamma = \pi/n$. It is important to have in mind that the limit for noncontextual correlations is given by $n- 2$, as is shown in Inequality~(\ref{eq:ncinequality}). Error bars are statistically calculated. It might seem surprising that the error does not increase with the number of settings. A quick calculation shows us, though, that for an ideal setup the variance of $\Omega$ is \[(\Delta\Omega)^2 = \frac{1}{N}n\sin^2(\pi/n) \le \frac{\pi^2}{nN},\] where $n$ is the number of settings and $N$ is the total number of countings. That is, the error \emph{decreases} with the number of settings.


\setlength{\tabcolsep}{1em}
\begin{table} [htpb]
\begin{tabular}{cccc}
\hline \hline
$n$ & $\Omega_{\rm exp}$ & $\Omega_{\rm exp}^{\rm bd}$ & $\Omega_{\rm max}^{\rm bd}$ \\
\hline
4 & $2.73\pm 0.02$ & 2.73 & 2.83 \\
6 & $4.90\pm 0.02$ & 5.11 & 5.20 \\
8 & $7.02\pm 0.02$ & 7.25 & 7.39 \\
10 & $8.80\pm 0.02$ & 9.25 & 9.51 \\
12 & $10.82\pm 0.02$ & 11.25 & 11.59 \\
14& $12.67\pm 0.02$ & 13.25 & 13.65 \\
\hline \hline
\end{tabular}
\caption{Values obtained for $\Omega$. $\Omega_{\rm exp}$ is associated with the raw experimental coincidence counts. $\Omega_{\rm exp}^{\rm bd}$ is the bound value obtained by the theory, when we use the angles implemented by the experiment, \textit{i.e.}, $\gamma_{\rm exp} = \kappa(x_i - x_s)$. $\Omega_{\rm max}^{\rm bd}$ is the maximal value allowed for quantum mechanics, which corresponds to using the angles  $\gamma = \pi/n$.} \label{tab1}
\end{table}



\begin{figure}[htpb]
\centering
\includegraphics[width=8.5cm]{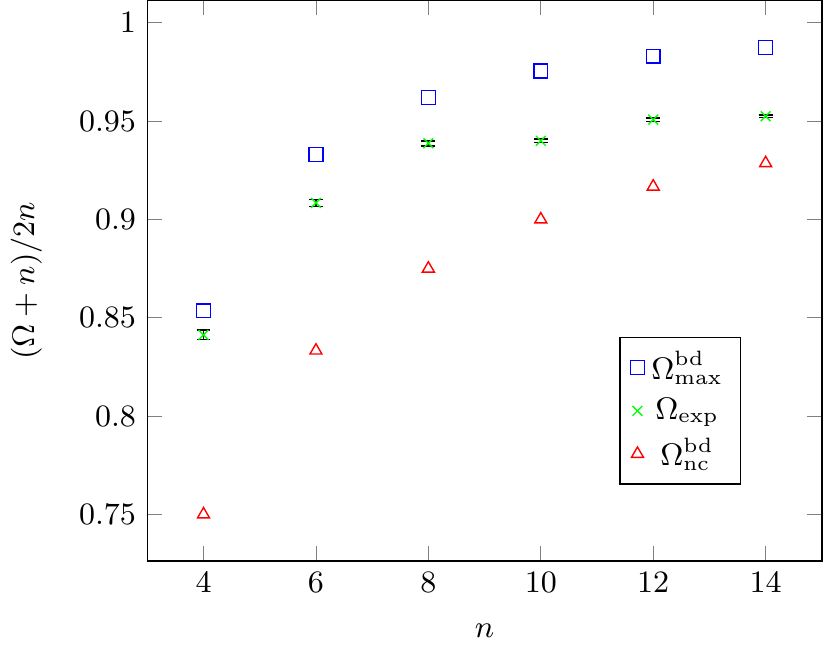}
\vspace{-0.5cm}
\caption{(Color online) Experimental results $\Omega_{\text{exp}}$, noncontextual bounds $\Omega^{\text{bd}}_{\text{nc}}$, and Tsirelson bounds $\Omega^{\text{bd}}_{\text{max}}$ for the $n$-cycle scenario. In all three cases we apply the transformation $\Omega \mapsto(\Omega+n)/2n$, so that the values can be interpreted as probabilities.}
\label{results}
\end{figure}

In Fig.~\ref{results} we plot the data from Table~\ref{tab1} transformed via $\Omega \mapsto (\Omega+n)/2n$. This was done so that we could interpret these values as the probability of success in the $n$-cycle game, which is a direct generalization of the prediction game proposed in \cite{liang11}. As it is characteristic for the families of $n$-cycle inequalities, the violations get smaller with increasing $n$. However, it is interesting to stress that our data show significant violations for all $n$ tested and if we were not limited by the positions of the detectors this setup could show significant violations above $n = 14$.


{\em Related work.---}Similar inequalities were measured with a non-maximally entangled state (and therefore non-maximal violation) for $n=6$ \cite{boschi97} and $n=42$ \cite{barbieri05} observables, with the goal of testing Hardy's ``nonlocality without inequalities'' \cite{hardy93}.


{\em Conclusions.---}In this work, we provide an experimental verification that a system of two photonic qubits, encoded in the transversal modes of entangled photons generated by the SPDC process, creates a probability distribution that is incompatible with any noncontextual hidden variable theory. Our experiment tests the violation of the noncontextuality inequalities for the $n$-cycle scenario for the case of even $n$, recently reported in~\cite{Araujo1}. It is important to stress that this is the first experimental violation of noncontextuality inequalities for a completely characterised infinite family of scenarios, and that now we have the experimental quantum violations for even $n$ up to $n=14$.

With a single two-particle conditional interference experiment, we are able to demonstrate the violation of six different noncontextuality inequalities with a number of settings ranging from $4$ to $14$. In this way, we have observed contextuality (in the sense of violation of the noncontextuality inequalities) in very good agreement with the predictions of quantum mechanics for six different scenarios.


\begin{acknowledgments}
 This work was supported by CNPq, CAPES, FAPEMIG, National Institute of Science and Technology in Quantum Information, the Science without Borders Program (Capes and CNPq, Brazil), and the Project No.\ FIS2011-29400 (MINECO, Spain).
\end{acknowledgments}



\end{document}